\documentclass[notitlepage,twocolumn,letterpaper,natbib,aps,prd,amsmath,amsfonts,nofootinbib,preprintnumbers,superscriptaddress,secnumarabic]{revtex4-1}
\pdfoutput=1
\usepackage{amssymb,amsmath,latexsym,mathrsfs}
\usepackage{url}
\usepackage{enumitem}
\usepackage{graphicx}
\usepackage{booktabs}
\usepackage[usenames,dvipsnames]{color}
\usepackage[breaklinks,colorlinks,urlcolor=magenta,citecolor=magenta,linkcolor=magenta]{hyperref}
\usepackage{multirow}
\usepackage{float}
\usepackage{cases}
\usepackage{blindtext}
\usepackage{pifont}
\usepackage{hhline}
\usepackage{xcolor}
\definecolor{linkcolor}{rgb}{0.0, 0.47, 0.75}
\definecolor{citecolor}{rgb}{1.0, 0.5, 0.0}
\hypersetup{
  linkcolor  = linkcolor,
  citecolor  = linkcolor,
  urlcolor   = linkcolor,
  colorlinks = true
}

\makeatletter
\def\maketitle{
\@author@finish
\title@column\titleblock@produce
\suppressfloats[t]}
\makeatother

\begin{document}

\preprint{CERN-TH-2023-167}
\vspace{5mm}

\title{Simulation-based inference for stochastic gravitational wave background data analysis}

\author{James Alvey}
\email{j.b.g.alvey@uva.nl}
\thanks{ORCID: \href{https://orcid.org/0000-0003-2020-0803}{0000-0003-2020-0803}}
\affiliation{GRAPPA Institute, Institute for Theoretical Physics Amsterdam,\\
University of Amsterdam, Science Park 904, 1098 XH Amsterdam, The Netherlands}

\author{Uddipta Bhardwaj}
\email{u.bhardwaj@uva.nl}
\thanks{ORCID: \href{https://orcid.org/0000-0003-1233-4174}{0000-0003-1233-4174}}
\affiliation{GRAPPA Institute, Anton Pannekoek Institute for Astronomy and Institute of High-Energy Physics,\\
University of Amsterdam, Science Park 904, 1098 XH Amsterdam, The Netherlands}

\author{Valerie Domcke}
\email{valerie.domcke@cern.ch}
\thanks{ORCID: \href{https://orcid.org/0000-0002-7208-4464}{0000-0002-7208-4464}}
\affiliation{Theoretical Physics Department, CERN, 1211 Geneva 23, Switzerland }

\author{Mauro Pieroni}
\email{mauro.pieroni@cern.ch}
\thanks{ORCID: \href{https://orcid.org/0000-0003-0665-266X}{0000-0003-0665-266X}}
\affiliation{Theoretical Physics Department, CERN, 1211 Geneva 23, Switzerland }

\author{Christoph Weniger}
\email{c.weniger@uva.nl}
\thanks{ORCID: \href{https://orcid.org/0000-0001-7579-8684}{0000-0001-7579-8684}}
\affiliation{GRAPPA Institute, Institute for Theoretical Physics Amsterdam,\\
University of Amsterdam, Science Park 904, 1098 XH Amsterdam, The Netherlands}

\begin{abstract}
\noindent The next generation of space- and ground-based facilities promise to reveal an entirely new picture of the gravitational wave sky: thousands of galactic and extragalactic binary signals, as well as stochastic gravitational wave backgrounds (SGWBs) of unresolved astrophysical and possibly cosmological signals. These will need to be disentangled to achieve the scientific goals of experiments such as LISA, Einstein Telescope, or Cosmic Explorer. We focus on one particular aspect of this challenge: reconstructing an SGWB from (mock) LISA data. We demonstrate that simulation-based inference (SBI) -- specifically truncated marginal neural ratio estimation (TMNRE) -- is a promising avenue to overcome some of the technical difficulties and compromises necessary when applying more traditional methods such as Monte Carlo Markov Chains (MCMC). To highlight this, we show that we can reproduce results from traditional methods both for a template-based and agnostic search for an SGWB. Moreover, as a demonstration of the rich potential of SBI, we consider the injection of a population of low signal-to-noise ratio supermassive black hole transient signals into the data. TMNRE can implicitly marginalize over this complicated parameter space, enabling us to directly and accurately reconstruct the stochastic (and instrumental noise) contributions. We publicly release our TMNRE implementation in the form of the code \texttt{saqqara}.

\vspace*{5pt} \noindent \textbf{\texttt{GitHub}}: The \texttt{saqqara} simulation and inference library is available \href{https://github.com/peregrine-gw/saqqara}{here} (\texttt{peregrine-gw/saqqara}). \\ 
In addition, the TMNRE implementation \texttt{swyft} is available \href{https://github.com/undark-lab/swyft}{here} (\texttt{undark-lab/swyft}).
\end{abstract}

\maketitle
\hypersetup{
  linkcolor  = linkcolor,
  citecolor  = linkcolor,
  urlcolor   = linkcolor
}

\section{Introduction}
\noindent With the detection of gravitational waves (GWs) by the LIGO-Virgo-KAGRA (LVK) collaboration~\cite{LIGOScientific:2016aoc}, and more recently by pulsar timing arrays (PTAs)~\cite{NANOGrav:2023gor,Antoniadis:2023rey,Xu:2023wog}, GW astronomy has now %well and truly 
entered the stage as a new player to explore our Universe. While existing GW observatories are noise dominated with transient signals that are relatively sparse and can be described by a few parameters, the situation will drastically change with the next generation of GW observatories. Both the space-based interferometer LISA~\cite{LISA:2017pwj} and the next-generation ground-based interferometers, such as the Einstein Telescope~\cite{Punturo:2010zz} or Cosmic Explorer~\cite{Reitze:2019iox}, are expected to see thousands of binary systems as well as the stochastic gravitational wave background (SGWB) of unresolved signals. This `orchestra' of overlapping signals poses a severe data analysis challenge to successful parameter reconstruction and component separation~\cite{Cornish:2005qw}.

Within this global analysis, the accurate reconstruction of an SGWB of cosmological origin is particularly challenging~\cite{Romano:2016dpx} -- but also offers a unique window to probe particle physics at energy scales far beyond the reach of colliders~\cite{Caprini:2018mtu}. Taking LISA as an example, the difficulties are quickly identified: we lack the possibility of carrying out cross-correlation (as in LVK or PTAs), and there are no perfect null-channels~\cite{Adams:2010vc, Muratore:2021uqj, Hartwig:2023pft} or the possibility to `shield' GWs, so the instrumental noise cannot be measured independently of a possible SGWB signal. Combined with the variety and complexity of possible particle physics models that lead to SGWBs, the accurate reconstruction of signal and noise parameters quickly becomes a highly challenging data analysis task.

A range of recent work has taken up this challenge using traditional inference techniques such as Markov Chain Monte Carlo (MCMC). The goal of the so-called LISA `global fit'~\cite{Cornish:2005qw, Vallisneri:2008ye, MockLISADataChallengeTaskForce:2009wir, Littenberg:2023xpl} is to simultaneously fit waveforms for different types of binaries
%reconstruct the parameters for different localized sources, 
%as well as 
and noise components. %This was demonstrated recently in Ref.~\cite{Littenberg:2023xpl} in the context of one of the LISA data challenges.
Focusing on LISA, and assuming that all individual above-threshold sources %have been accurately reconstructed and 
are removed, several approaches have demonstrated the possibility of achieving the simultaneous reconstruction of SGWB and noise. These rely on templates either for the noise~\cite{Karnesis:2019mph, Caprini:2019pxz, Pieroni:2020rob, Flauger:2020qyi}, the SGWB signal~\cite{Baghi:2023qnq, Muratore:2023gxh}, or both~\cite{Boileau:2020rpg, Hartwig:2023pft}. The overall challenge is the sheer dimensionality of the problem (in principle, there will likely be at least $\sim 10^5$ parameters in the full problem, see, e.g., Refs.~\cite{Cornish:2005qw,Littenberg:2023xpl}), and the high precision reconstruction required to extract an SGWB signal. In a very broad sense, the goal of our work is to argue that simulation-based inference (SBI) techniques may be a promising path toward mitigating these traditionally conflicting goals of precision and scale. Moreover, an SBI pipeline could offer an independent cross-check to validate the results obtained with traditional methods.

SBI techniques (see e.g. Ref.~\cite{Cranmer:2019eaq} for a review) have recently undergone a significant uptick in popularity as we approach a new era of big data analysis challenges. 
In contrast to stochastic-sampling approaches such as MCMC or nested sampling, SBI algorithms look to solve the Bayesian inference problem of reconstructing the posterior distribution, $p(\boldsymbol{\theta} | \boldsymbol{x})$, without requiring an explicit expression for the likelihood $p(\boldsymbol{x} | \boldsymbol{\theta})$ (although there are additional, independent benefits including e.g. amortisation, scalability, and simulation efficiency).
Instead, the likelihood distribution is sampled implicitly via some stochastic forward simulator that generates data $\boldsymbol{x}$ from parameters $\boldsymbol{\theta}$. This fundamentally shifts the focus from building a statistical model to developing a realistic computational forward model for the data, including e.g. all relevant instrumental and physical effects. SBI methods have now been shown to perform inference to the level of a full likelihood-based approach
in several astrophysical and cosmological settings, including GW data analysis, see e.g.~\cite{Alvey:2023naa,Alvey:2023pkx,Bhardwaj:2023xph,Dax:2022pxd,Dax:2021tsq,Montel:2022fhv,AnauMontel:2022ppb,Dimitriou:2022cvc,Karchev:2022xyn,Makinen:2021nly}. Whilst several SBI algorithms exist, %there are a range of SBI algorithms, 
see Refs.~\cite{Papamakarios:2016aaa,Tejero-cantero:2020aaa,Zeghal:2022aaa,Papamakarios:2016aaa,Alsing:2019xrx,Lin:2022ayr,Rozet:2022aaa}, in this work, we will focus on the application of  (truncated marginal) neural ratio estimation (TMNRE)~\cite{Miller:2022shs}, implemented within the code \texttt{swyft}~\cite{Miller:2021hys}.

Several crucial benefits suggest the TMNRE algorithm could be an ideal tool for LISA SGWB data analysis. Firstly, the truncation aspect (which makes TMNRE a sequential SBI algorithm) allows us to effectively ``zoom-in" on the relevant regions of parameter space for a given observation (see Refs.~\cite{Miller:2021hys,Bhardwaj:2023xph} for details on this procedure). In a variety of cases, such as for cosmic microwave background data~\cite{Cole:2021gwr}, strong lensing image analysis~\cite{Montel:2022fhv}, and GWs from compact binary coalescences~\cite{Bhardwaj:2023xph,Alvey:2023naa}, this makes TMNRE extremely simulation efficient compared to both traditional methods and non-sequential SBI algorithms. Indeed, Ref.~\cite{Bhardwaj:2023xph} demonstrated that analysing LIGO-type binary black hole mergers with TMNRE requires 98\% fewer waveform evaluations than the currently adopted nested sampling approach. Secondly, the TMNRE algorithm can target specific parts of the model while implicitly marginalising over all other components~\cite{AnauMontel:2023stj}. We will highlight this property in our analysis and demonstrate that we can directly analyse only the noise and SGWB components, properly marginalised over additional transient sources. Third, realistic LISA data will contain numerous GW signals (binaries and SGWBs) and instrumental noise.
%The forward modelling of GW waveforms, known individual noise components, and instrument response functions is well established, whereas the complexity of the problem could prohibit an explicit, exact expression for the likelihood $p(\boldsymbol{x} | \boldsymbol{\theta})$ rendering the inverse problem of parameter estimation very difficult. 
The methodologies and pipelines for the forward modelling of GW waveforms, known individual noise components, and instrument response functions are, up to technical refinements, well established for LISA. Conversely, the complexity of the problem could prohibit an explicit, exact expression for these marginalised likelihoods, rendering the inverse problem of parameter estimation potentially very costly, since one would have to work with the full likelihood. 
The implicit likelihood benefits that are at the core of SBI can circumvent these difficulties.

Ultimately, our proposed use-case for this algorithm has %exactly 
the same spirit in mind as the `global fit'~\cite{Littenberg:2023xpl}: separating the multiple components in a LISA data stream.
As such, if possible, it is useful to split these into distinct analysis `blocks' to combat the dimensionality issues and consistently pass the subsequent inference results around the full model. Here, we illustrate how one could do this for the `block' containing the SGWB and noise components, accounting for, e.g., the presence of transient sources. In this regard, our analysis should be seen as a first step, dealing with a setup that, in many ways, is simplified compared to the data analysis challenge that LISA will face. However, this proof of principle and verification against other methods is a key step to unlocking the potential of SBI for GW data analysis.

%\vspace*{8pt}
\noindent \textbf{Code:} Along with the results presented here, we also provide an extendable public code which can be found \href{https://github.com/peregrine-gw/saqqara}{here} (\texttt{peregrine-gw/saqqara}).

\section{Analysis Setup and Data Generation}
\noindent To explore the ability of SBI -- and more specifically TMNRE -- to address the challenges of SGWB analysis, we set up several %a number of 
case studies to analyse. These are broadly similar to those presented in Ref.~\cite{Flauger:2020qyi} and cover both the recovery of a signal given a parameterised template, as well as the agnostic fitting of an unknown signal. We then extend the analysis to include transient signals for a mock population of supermassive black holes to investigate the implications at the level of parameter estimation. Given this, there are several %a number of 
technical components to setting up the analysis: a model for the instrument noise in LISA; characterisation of the SGWB templates; explanation of the transient setup; and data generation. 

Considering the instrument noise first, currently, the knowledge of the LISA noise comes from LISA Pathfinder (LPF)~\cite{Castelli:2020zro}, which tested the purity of free-fall for the Test Masses (TM), and from on-ground experiments. A two-parameter noise model, specified in terms of low-frequency TM noise and high-frequency Optical Metrology System (OMS) noise, defines a reasonable approximation of the LISA noise~\cite{LISA:2017pwj, Babak:2021mhe}. Each noise component depends quadratically on a parameter (referred to as $A$ and $P$, respectively, and with fiducial values $A=3$, $P=15$), which controls its amplitude. For more details on the LISA noise model and the measurements LISA will perform, see Sec.~I in the \textit{Supplemental Material}. Consistent with Refs.~\cite{Caprini:2019pxz, Pieroni:2020rob, Flauger:2020qyi, Babak:2021mhe, Hartwig:2023pft}, in the present work, we make the somewhat strong assumption that the noise {\it shapes} are perfectly known while we allow for the amplitude to vary %freely upon including 
within a wide, uniform prior for the noise parameters (assumed to be positive) centered around the fiducial values. However, we stress that tackling realistic LISA data analysis will require more complex noise modelling, which we leave to future work.

As far as the SGWB signal itself is concerned, we consider two types of templates: a power law (PL) specified by a tilt ($\gamma$) and (log) amplitude ($\alpha$); and a more agnostic form %which is 
defined by a (log) amplitude in the first bin $\alpha_1$ and a sequence of slopes $\gamma_j$ for $j = 1, \cdots, N_\mathrm{bins}$, where $N_\mathrm{bins}$ is the number of equally spaced logarithmic bins that the template is split into. In this work, we consider $N_\text{bins} \leq 10$, though we note that Ref.~\cite{Dimitriou:2023knw} recently demonstrated the possibility of scaling to a larger ($\sim 20$) number of bins. More concretely, we specify the templates (written in terms of the GW energy density $\Omega_\mathrm{GW}h^2$) as
\begin{align}
    \textbf{power law:}\quad \Omega_{\rm GW}(f)h^2  & = 10^{\alpha} \left(\frac{f}{\sqrt{f_{\mathrm{min}}f_{\mathrm{max}}}}\right)^{\gamma} \nonumber \\
   \textbf{agnostic:}\quad\Omega_{\rm GW}(f)h^2 & = \sum_{i = 1}^{N_\mathrm{bins}} 10^{\alpha_i} \left(\frac{f}{\sqrt{f_{\mathrm{min}, i}f_{\mathrm{max}, i}}}\right)^{\gamma_i} \nonumber \\ 
   & \times \Theta(f - f_{\mathrm{min}, i}) \Theta(f_{\mathrm{max}, i} - f) \nonumber
\end{align}
where $f_\mathrm{min} = 10^{-4}\,\mathrm{Hz}$, $f_\mathrm{max} = 5 \times 10^{-2}\,\mathrm{Hz}$, $f_\mathrm{min/max, i}$ are the boundaries of each of the bins, $\Theta$ denotes the Heaviside function, and $\alpha_i$ and $\gamma_i$ are the amplitude and tilt in each bin. In the analysis, we will vary each of the corresponding parameters uniformly in prior ranges specified in Tab.~\ref{tab:priors}. Imposing continuity, this fixes all the $\alpha_i$ for $i \geq 2$. 

\begin{table}[t]
\centering
\begin{tabular}{@{}ll@{}}
\hline
\textbf{Parameter} & \textbf{Prior Choice} \\ \hline
(Log) Amplitudes, $\alpha$, $\alpha_1$ & $\mathrm{U}(-20, -5)$ \\
Tilts, $\gamma$, $\gamma_i$ & $\mathrm{U}(-10, 10)$ \\ 
TM Noise, $A$ & $\mathrm{U}(0, 6)$ \\
OMS Noise, $P$ & $\mathrm{U}(0, 30)$ \\ \hline
\end{tabular}
\caption{Prior choices for the case studies and analyses presented in this work for the (dimensionless) amplitudes, tilts, and instrumental noise parameters.}
\label{tab:priors}
\end{table}

Beyond this, there are several %a number of general 
assumptions that we make in our data generation. First, we work with a single time-delay interferometry (TDI) channel.  Moreover, we model the time domain data $d(t)$ as a superposition of one (or more) signal component(s) $s_c(t)$, and detector noise $n(t)$ as 
$d(t) = n(t) + \sum_c s_c(t)$. We assume stationarity in the time-domain data,\footnote{The assumption of stationarity does not apply to transient sources, or if so, only statistically.} which implies vanishing correlations between different frequencies in the Fourier domain. We consider mock data corresponding to an observation time $T_{d}$ of 12 days.\footnote{%We note that t
This corresponds to about 1/100th of the planned LISA observation time, however, it suffices to demonstrate our main points regarding the statistical flexibility and precision agreement of our algorithm. In addition, though, we explicitly tested the pipeline with $\Delta f = 10^{-5} \,\mathrm{Hz}$, or about 115 days of data split into 100 segments, and achieved similarly good agreement with MCMC. We leave the task of optimally scaling this up (e.g. via a similar coarse-graining scheme to Ref.~\cite{Flauger:2020qyi}) to the full length of the LISA data -- including importantly non-stationary noise components -- for future work.} For data compression, we divide it into $N_d = 100$ data segments of duration $T_s \equiv T_{d}/N_d$ each. In this scheme, the frequency resolution in each data segment is $\Delta f = 1/T_{s} = 10^{-4}$~Hz, and we denote with $\tilde{d}_s(f_\textmd{k})$ the frequency-domain data for each segment $s$ and (discrete) frequency $f_\textmd{k}$. We also assume signal and noise to be Gaussian distributions with zero mean and variances based on their respective power spectral densities (PSDs). Under these assumptions, we generate $N_d$ statistical realizations of the SGWB signal and noise. In the final analysis, %that we consider, 
we also investigate the implications of introducing a population of transient signals. To do so, we introduce a probability $p$ that a given data segment contains a transient %has a transient in it 
(this could easily be extended to include multiple transients). For each data segment, we inject a mock supermassive-black hole waveform with probability $p$.\footnote{Specifically we take $p=0.05$, which on average would introduce 5 sources. %Assuming 11 parameters per source, this would correspond to 55 additional parameters to constrain (in this particular example).}
} We use the \texttt{IMRPhenomXAS} waveform template implemented in the \texttt{jax}-based \texttt{ripple} package~\cite{Edwards:2023sak} to generate the frequency-domain strains for this signal component. For the explicit choices of the population parameters see the \textit{Supplemental Material}.

Finally, there are several details that are relevant for constructing the comparison to the MCMC method. Specifically, following the approach introduced in~\cite{Caprini:2019pxz, Flauger:2020qyi}, we define averaged data $\bar{D}_{\textmd{k}} \equiv \sum_s \tilde{d}_s(f_\textmd{k}) \tilde{d}^{*}_s(f_\textmd{k}) / N_d$ and down-sample it through coarse-graining. This yields a new (binned) dataset $D_{\hat{k}}$, where $\hat{k}$ covers a sparser set of frequencies $f_{\hat{k}}$, and comes with weights $w_{\hat{k}}$~\cite{Caprini:2019pxz, Flauger:2020qyi}. An unbiased (log-) likelihood for the compressed dataset can be built, e.g., as a mixture of a Gaussian and a log-normal component~\cite{Flauger:2020qyi}. For the explicit expression see Sec.~I in the \textit{Supplemental Material}. To sample the parameter space, we use the \texttt{emcee} sampler described in Ref.~\cite{Foreman-Mackey:2012any}. Given this setup, we can define the four benchmark analyses that we present the results for below:

\vspace*{-6pt}
\begin{itemize}\setlength{\itemsep}{-4pt}
    \item[\textbf{C1:}] PL template and LISA noise
    \item[\textbf{C2:}] Agnostic template with $5$ bins and LISA noise
    \item[\textbf{C3:}] Agnostic template with $10$ bins and LISA noise
    \item[\textbf{C4:}] PL template, LISA noise, and additional transients
\end{itemize}
\vspace*{-6pt}

\section{Simulation-based Inference Framework}
\noindent The implementation (as presented in \texttt{saqqara})% the \texttt{saqqara} library) 
of the TMNRE~\cite{Miller:2021hys} algorithm in the context of SGWB recovery is one of the key results of this work. As such, we devote this short subsection to a description of some of the specific design choices relevant to SGWB analysis. We do not cover detailed explanations of the algorithm but instead refer the reader to, e.g., Ref.~\cite{Miller:2021hys} for the initial presentation of TMNRE. In addition, we point the reader to Sec.~2 of Ref.~\cite{Bhardwaj:2023xph} for a detailed description of the application of TMNRE to GWs from compact binary coalescence sources, outlining a lot of the logic we also follow in this work. Finally, the implementation of the autoregressive ratio estimation used to explore the parameter space in the final stage of inference is described in Ref.~\cite{AnauMontel:2023stj}.

\begin{figure*}[t]
    \centering
    \includegraphics[width=0.8\linewidth]{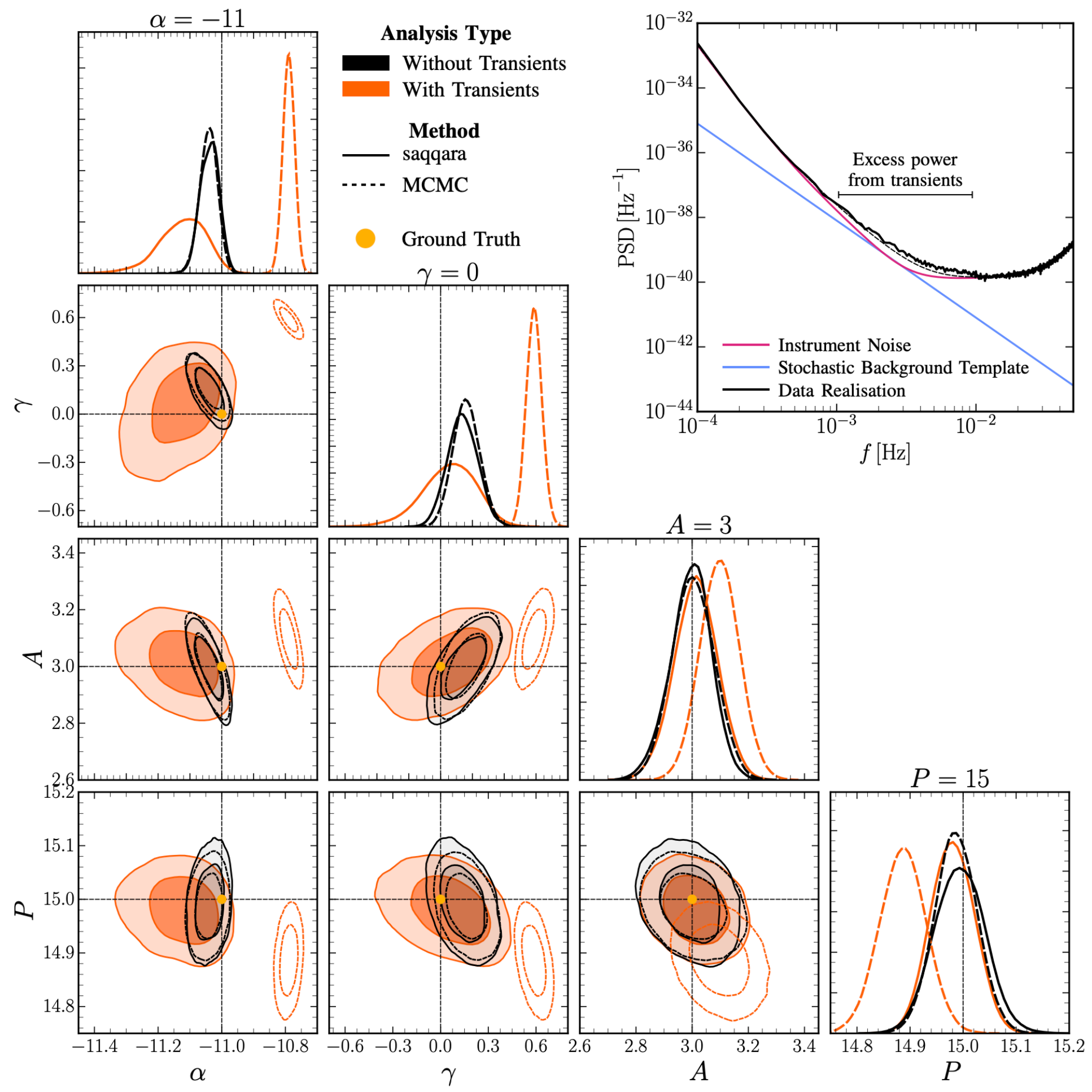}
    \caption{Analysis results for the case studies \textbf{C1} and \textbf{C4}. \textit{Main Plot.} Corner plot highlighting the two analysis results for case studies \textbf{C1} -- which corresponds to the PL template without additional transients (shown in black solid and dashed contours) -- and \textbf{C4} -- where the transients are now present (shown in orange solid and dashed contours). The true injected values are highlighted by the dashed black horizontal and vertical lines, and by the yellow markers. \textit{Upper inset.} Illustration of the explicit data realisation (black line) for the case study \textbf{C4} along with the injected instrumental noise (pink line), stochastic background signal (blue line), and their sum (dashed black curve).}
    \label{fig:powerlaw-main}
\end{figure*}

To understand where the design choices are made, it is useful to think of the ratio estimation step in TMNRE in two parts: compression, and ratio estimation. In particular, to implement the TMNRE algorithm, one must design a network architecture that can take in data $\boldsymbol{x}$ and parameters $\boldsymbol{\theta}$ and estimate the ratio $r(\boldsymbol{x}, \boldsymbol{\theta}) = p(\boldsymbol{x} | \boldsymbol{\theta}) / p(\boldsymbol{x})$, where $p(\boldsymbol{x})$ is the (Bayesian) model evidence. In practice, $\boldsymbol{x}$ is usually (or at least can be) a high-dimensional and complicated data structure, so we normally first define a compression network that compresses $x$ to some lower-dimensional summary $s(\boldsymbol{x})$. This summary $s(\boldsymbol{x})$ is then combined with $\boldsymbol{\theta}$ and inputted into the standard ratio estimators implemented within \texttt{swyft}~\cite{Miller:2022shs}. Importantly, both the summary and the ratio estimators are optimised simultaneously, resulting in %some sort of 
an automatically learned summary statistic. As such, for various applications, the main novelty in the implementation lies in the design of the compression network that produces the summary $s(\boldsymbol{x})$. In the current context, we make use of the following structure. Starting with SGWB data $\boldsymbol{x}$ that is a time-averaged sequence of frequency bins, as in Fig.~\ref{fig:powerlaw-main}, we first perform a normalisation step by taking the natural logarithm and then applying an online normalising layer~\cite{Miller:2022shs}. This is purely for performance reasons in the sense that the network optimisation proceeds more robustly on normalised data. In the next stage, we do consider the structure of the data, however, and utilise an architecture similar to that described in Ref.~\cite{Bhardwaj:2023xph}. In particular, we apply a set of 1-dimensional convolutional layers, organised into a UNet architecture~\cite{Ronneberger:2015aaa}. The motivation for this is to allow both the local and global sharing of information across the various frequency bins as we look to learn the SGWB and noise templates. Scalability to more complex situations also guided this part of our architecture, allowing for direct application to either multiple channels or correlated noise without any modifications. Finally, we added a simple linear compression network to summarise this information into a lower dimensional vector that can then be combined with the parameters $\boldsymbol{\theta}$. The remainder of the network consists of the standard ratio estimators\footnote{In particular, we can either estimate the individual 1-dimensional marginals, higher dimensional marginals, or an autoregressive estimator. All of these are implemented within \texttt{saqqara}.} implemented in \texttt{swyft} and described in Refs.~\cite{Miller:2022shs,AnauMontel:2023stj}, optimised on the standard binary cross-entropy loss relevant to neural ratio estimation~\cite{Miller:2022shs,Rozet:2022aaa,Durkan:2020aaa,Hermans:2019aaa}. As a reference, we provide the various numerical settings choices for the algorithm both in the \texttt{saqqara} release, as well as in the \emph{Supplemental Material}, where we also discuss the computational performance.

When designing this architecture, one motivation was to ensure its scalability to more complex situations. For example, with this implementation, we can directly apply this to a LISA data structure containing either multiple channels or correlated noise without any modifications. In addition, looking towards analyses carried out in the time-frequency domain, aside from possibly slight modifications to the compression network, the \emph{entirety} of the rest of the pipeline can remain unchanged. This opens up the possibility of applying this algorithm directly to, e.g.,~the separation of several SGWB components, more complex noise models including non-stationary noise scenarios, or varying detector configurations, which is something we aim to do in future work.

\section{Results and Discussion}
\begin{figure*}[t]
    \centering
    \includegraphics[width=\linewidth]{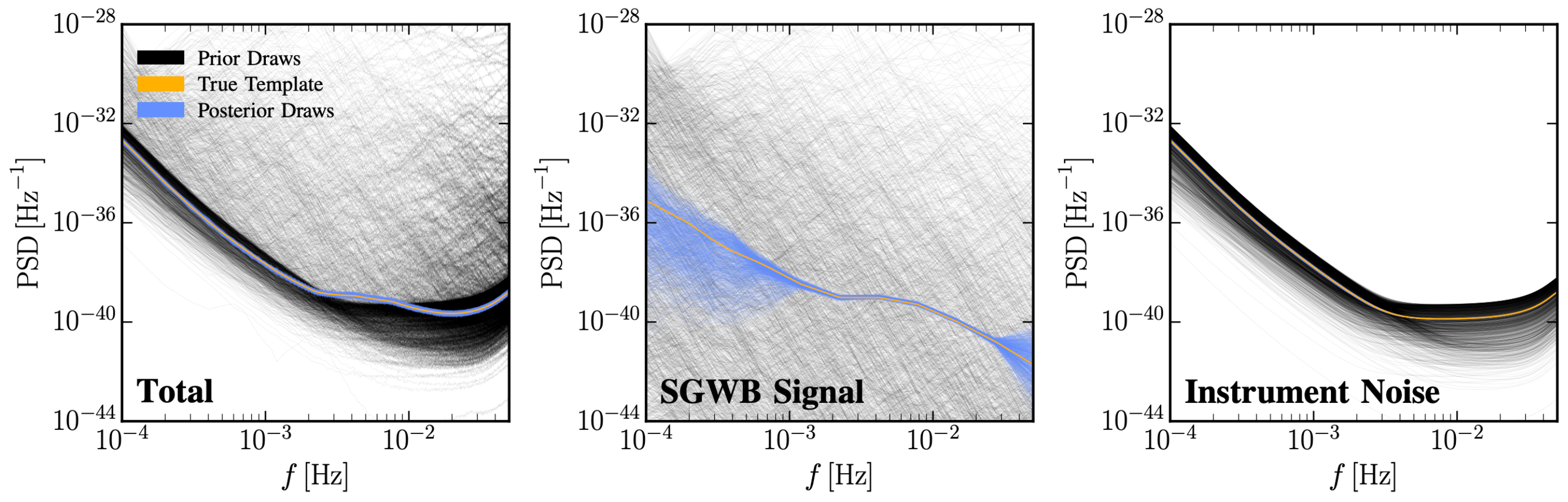}
    \caption{Reconstruction of stochastic background and instrumental noise components for an agnostic template with ten signal bins. In all panels, draws from the initial Bayesian prior are shown in black, whereas the resulting posterior draws are shown in blue. The injected signal is shown in yellow. \textit{Left panel.} Reconstruction of the total power-spectral density (PSD). \textit{Middle panel.} Reconstruction of the individual stochastic background component. \textit{Right panel.} Reconstruction of the instrumental noise contribution.}
    \label{fig:agnostic-main}
\end{figure*}

\noindent All of the key results for this work are in the context of the case studies described above and are summarised in Fig.~\ref{fig:powerlaw-main} and Fig.~\ref{fig:agnostic-main}. In brief, they highlight two key points: firstly, %we are able to use SBI to reproduce 
SBI techniques reproduce the results from traditional sampling methods (MCMC in this case). This is true both in the case of a PL template, as well as a more agnostic fit. Secondly, when we introduce the additional complexity from transient sources, our SBI method still produces unbiased %and precise 
posterior distributions without any modification. %, in contrast to a direct likelihood based approach.

In more detail, we will first discuss the agreement with traditional methods. To do so, we take case studies \textbf{C1}, \textbf{C2}, and \textbf{C3}. The first of these is shown by the black solid and dashed contours in Fig.~\ref{fig:powerlaw-main}. These highlight very clearly the fact that we can accurately reproduce the unbiased and accurate posteriors obtained from a traditional likelihood-based analysis for a PL template. Furthermore, in Fig.~\ref{fig:agnostic-main}, we illustrate our ability to constrain a more agnostic template (in this case with 10 signal bins, although the result with 5 bins is shown in the \textit{Supplemental Material}). The three panels show the relative constraining power of the analysis via the posterior draws (blue lines) compared to prior draws (black lines) for the total signal, as well as the separate SGWB signal and instrumental noise contribution. For comparison, the injected signal is shown in yellow across all panels. We see that we reproduce several %a number of 
desirable characteristics in this agnostic case, for example, the fact that we obtain tight constraints on the SGWB signal when it dominates over the instrumental contribution, and wider constraints in the opposite case, i.e.,~outside the frequency band in which the instrument is most sensitive. The specific agreement with MCMC at the level of the posterior distribution over parameters for the five bin case is provided in the \textit{Supplemental Material}. Again, this highlights our precise agreement with traditional sampling-based methods.

The second key result is presented via the orange contours in Fig.~\ref{fig:powerlaw-main}. This is the result for the case study \textbf{C4} described above, where we introduce a population of supermassive black hole transients into the data. At the level of data realisations, we can easily understand the effect of this by looking at the upper inset in Fig.~\ref{fig:powerlaw-main}. In particular, the additional transients lead to a relative (and realisation dependent) excess in the mid-frequency region compared to the noise templates alone. It is important to note that this excess is not distributed in the same way as the instrumental noise and SGWB contributions (which are distributed as coloured zero mean Gaussian noise). This means that the impact of the transients cannot be simply accounted for in a likelihood-based approach unless each signal is individually analysed and the full parameter space for transient signals is sampled within the MCMC. The cost of this is significant, however, since the relative dimensionality of the problem would then increase by a huge margin in line with the number of signals (hundreds) multiplied by the number of signal parameters (tens). If we look to avoid this and take the naive approach by analysing the data using the same likelihood model as in \textbf{C1}-\textbf{C3}, we obtain the dashed orange contours. Not unexpectedly, these are significantly biased compared to the true injection value shown by the yellow dot. We do emphasise that this is not necessarily an intrinsic limitation of the MCMC approach, since we knew that the likelihood was incorrect. What this result emphasises, however, is that if one wants to use traditional sampling methods, there is a crucial need for either \emph{(a)} clean data with all transients removed, or \emph{(b)} a higher-dimensional likelihood fitted to each transient, at the cost of significant sampling time. 

At this point, we present the approach using our SBI algorithm, which looks to provide a compromise between these two options. Essentially, SBI allows us to implicitly marginalise over the complicated parameter space introduced by the transients by learning the effective marginalised likelihood-to-evidence ratio. This only depends on the SGWB and instrumental noise parameters and can be directly sampled. The results of this process are shown by the solid orange contours in Fig.~\ref{fig:powerlaw-main}. Crucially, we see that we are able to provide an unbiased and precise posterior that correctly accounts for the presence of the transients via a slight, but noticeable broadening of the SGWB parameter ($\alpha$ and $\gamma$) posteriors. Furthermore, we see that we are actually able to obtain identical constraints on the noise parameters as compared to the \textbf{C1} case. There are two reasons for this: firstly, in our setup, the transients mainly affect the mid-frequency region, so the noise parameters can be constrained to the same precision from the low- and high-frequency data, and secondly, we used the same data realisation for the SGWB and noise components. In addition to this, we carry out standard posterior coverage tests~\cite{Hermans:2021aaa} for this case study, which are provided in the \textit{Supplemental Material}. We find that for all parameters, our posteriors are extremely well calibrated, adding strength to the claim that we correctly marginalise over the additional transient components. This is a key result of this work and is the connection point to something resembling the LISA global fit challenge. In particular, this shows that SBI provides the possibility to directly analyse the SGWB or instrumental noise component of a LISA data stream without necessarily removing all transient artifacts. Instead, they could be simply included in the data generation, and implicitly marginalised over using the procedure presented in this work.

\section{Conclusions and Outlook}
\noindent This work is a first step in demonstrating the potential of  SBI techniques in addressing the data analysis challenges of LISA. Specifically, we showed in the case studies \textbf{C1}-\textbf{C3} that posteriors computed with SBI match the ones obtained with traditional likelihood-based methods. The final study (\textbf{C4}) shows that these results are robust even in the presence of additional transients (when included in the training data), demonstrating the potential of directly estimating marginal posteriors, which poses a serious challenge in the MCMC approach. While we choose a specific class of transients, we anticipate this result to hold for different signals (or noise features), too. More investigations are required to test the robustness and limitations of the methodology. This characteristic of SBI opens up the intriguing possibility of integrating similar techniques in (or using them as independent cross-checks for some parts of) the LISA global fit pipeline.

Building on this, future work will need to address (i) the inclusion of multiple data channels (nominally the standard A, E, and T basis that is well-established in the literature), (ii) more realistic source modelling, providing either a full range of possible sources or a representative and realistic output of an initial analysis where sources are fully or partially removed, as well as robustness tests in view of limited source knowledge and (iii) instrumental noise that is less well-calibrated, non-stationary, correlated, and/or contains more relevant noise components. We stress that none of these strike us as fundamental obstacles within the infrastructure presented here and that the simplifying assumptions taken in this work only served as a starting point in the development of these techniques.
In addition, we envisage applications of SBI techniques to perform parameter estimations in other blocks of the LISA analysis problem, maintaining the ability to correctly marginalise over all other parameters.  In this spirit, we hope that the release of our public code will trigger some of these developments. 

\vspace*{12pt}
\noindent \emph{Note added:} In parallel to this work, neural posterior estimation (as opposed to neural ratio estimation used here) was investigated in Ref.~\cite{Dimitriou:2023knw}, demonstrating in particular the feasibility of efficiently reconstructing signals using the full 3 years of LISA data and 3 data channels. They follow the same simplifying assumptions on the noise model as employed in this work.

\vspace*{-12pt}
\section*{Acknowledgements}
\vspace{-8pt}
\noindent We are grateful for fruitful discussions at the TH Institute on SGWB data analysis organized at CERN, in particular with A.~Dimitriou, D.~Figueroa, and B.~Zaldivar. 
This work is part of the project CORTEX (NWA.1160.18.316) of the research programme NWA-ORC which is (partly) financed by the Dutch Research Council (NWO).
Additionally, CW acknowledges funding from the European Research Council (ERC) under the European Union’s Horizon 2020 research and innovation programme (Grant agreement No. 864035). UB is supported through the CORTEX project of the NWA-ORC with project number NWA.1160.18.316 which is partly financed by the Dutch Research Council (NWO). JA is supported through the research program ``The Hidden Universe of Weakly Interacting Particles" with project number 680.92.18.03 (NWO Vrije Programma), which is partly financed by the Nederlandse Organisatie voor Wetenschappelijk Onderzoek (Dutch Research Council). JA acknowledges the hospitality of CERN TH, where this work was initiated. JA and MP acknowledge the hospitality of Imperial College London, which provided office space during some parts of this project.

\bibliography{biblio}

\appendix
\onecolumngrid

\section{LISA measurements}\label{app:measurements}

\noindent This section summarizes the formalism to characterize signal and noise in LISA. Following Ref.~\cite{Hartwig:2023pft}, we start with a brief description of single link GW measurements, proceed by discussing the noise PSDs, and conclude by introducing Time Delay Interferometry (TDI) variables. As discussed in the main body of this work, the data $d(t)$ contains a linear superposition of signal and noise, and has a Fourier transform\footnote{We are assuming $d(t)$ is a continuous function of time, which is not strictly correct since, in reality, data sampling occurs at a finite rate (and typically is impacted further by some downsampling procedure). On the other hand, since these effects will only affect the high-frequency end of the spectrum ($f \gtrsim 1$~Hz), which we do not include in the analysis, we can safely ignore this effect.} which reads
\begin{equation}
	\tilde{d}(f) = \int_{-T/2}^{T/2} \textrm{e}^{2 \pi i f t } \; d(t) \; \textrm{d} t \; ,
\end{equation}
where $T$ is the observation time. Assuming these components are uncorrelated -- which implies we can treat them independently -- and stationary -- which implies vanishing correlations between different frequencies -- we obtain
\begin{equation}
    \label{eq:data_correlation}
    \langle \tilde{d}(f)  \tilde{d}^*(f')  \rangle = \frac{1}{2} \delta(f- f')\left[ S^{\mathrm{N}}(f) + S^{\mathrm{GW}}(f)\right] = \frac{1}{2} \delta(f- f')\left[ S^{\mathrm{N}}(f) + \sum_\lambda \mathcal{R}_\lambda (f) P^{\lambda}_h (f) \right] \; ,
\end{equation}
where $S^{\mathrm{N}}(f)$, $S^{\mathrm{GW}}(f)$ are the noise and signal PSDs. These are real, positive, and even functions of $f$. In the second equality, we have further expanded $S^{\mathrm{GW}}(f)$ in terms of $\mathcal{R}_\lambda$, the sky-averaged LISA response function, which projects the GW PSD $P^{\lambda}_h$ (with $\lambda$ running over the two GW polarizations), onto the data. For this purpose, we have assumed $P^{\lambda}_h$ to be homogeneous, isotropic, and diagonal in the GW polarization basis (for details, see, e.g. Refs.~\cite{Bartolo:2018qqn,Domcke:2019zls,Flauger:2020qyi, Hartwig:2023pft}). For completeness, we recall that, given the intensity $I(f) \equiv \sum_{\lambda} P^{\lambda}_h /2$, the SGWB energy density reads:
\begin{equation}
    \Omega_{\rm GW} h^2 \equiv  \frac{4 \pi^2  }{3 (H_0/h)^2} f^3 I(f) \; ,
\end{equation}
where $H_0 \approx 3.24 \times 10^{-18} h$ Hz is the Hubble constant today and its dimensionless value~\cite{Planck:2018vyg} is $h = 0.6766 \pm 0.0042$. Since planar interferometers like LISA are not sensitive to chirality\footnote{Due to symmetry, left and right-hand polarized GW waves coming from opposite directions would induce the same effect in a planar interferometer. Possible ways to break this degeneracy and measure chirality include, e.g., correlating the signal measured by non-coplanar interferometers~\cite{Seto:2007tn, Crowder:2012ik, Orlando:2020oko} or using the dipole induced by the motion of the detector with respect to the SGWB frame~\cite{Seto:2006hf, Seto:2006dz, Domcke:2019zls}.}, $\mathcal{R}_{\rm L} = \mathcal{R}_{\rm R}$ (with L, R, denoting the two GW helicities), and thus, the data only depends on $I(f)$.

For the transient signals, we generate a database of 2 million waveforms evaluated on the same frequency grid as in the main analyses.\footnote{Assuming each merger to occur within a data segment, the observation time defines, beyond the frequency resolution, the minimal frequency at which the system emits in that segment. In practice, this would correspond to cutting each of the transients at some minimal frequency, such that the time for the evolution from this frequency to the mergers is $\leq T_c$. For the sake of simplicity, we ignore this effect, which, in some sense, is only an artifact of the short observation period used in this analysis.}
Each waveform has parameters sampled uniformly from the priors: $M_c \in [8, 9]\times 10^5\,M_\odot$ (chirp mass), $\eta \in [0.25, 1]$ (mass ratio), $\chi_1, \chi_2 \in [-1, 1]$ (dimensionless spins), $d_L \in [5, 10] \times 10^4 \, \mathrm{Mpc}$ (luminosity distance), $t_c = 0\,\mathrm{s}$ (time of coalescence), and $\phi_c = 0$ (phase).  We then rescale the strain amplitude of the transient by $10^{-3}$ to make them behave like a population of sub-threshold sources. While this setup is not fully realistic, it provides a case study to show that these injections are sufficient to bias the MCMC approach.

\subsection{Single link signal response and noise PSDs.}
\noindent Following the notation of Ref.~\cite{Hartwig:2023pft}, the fractional frequency shift $\eta_{ij}^{\mathrm{GW}}(t)$ induced by a GW perturbing the path of a photon released at time $t - L_{ij}$ from an emitter located at $\vec{x}_i$, to a receiver $\vec{x}_j$ at time $t$, (with $L_{ij} \equiv |\vec{x}_i - \vec{x}_j|$), reads
	\begin{equation}
		\eta_{ij}^{\mathrm{GW}}(t) =  i \int_{-\infty}^{\infty} \textrm{d} f\, \frac{f}{f_{ij}} \, \textrm{e}^{2\pi i f(t - L_{ij})}
		\int \textrm{d} \Omega_{\hat{k}} \, \textrm{e}^{-2\pi i f \hat{k}\cdot \vec x_i } \sum_\lambda \xi^\lambda_{ij}(f, \hat k) \, \tilde{h}_\lambda(f,\hat{k})  \; ,
		\label{eq:etaij_final}
	\end{equation}
where we have expanded the GW in plane waves, with $\vec{k}$ being the GW momentum, $\Omega_{\hat{k}}$ denoting the solid angle, $\tilde{h}_\lambda(f,\hat{k})$ being the coefficients of the expansion, and $\lambda$ running over the two GW polarizations. We have also introduced the characteristic frequencies $f_{ij} \equiv (2 \pi L_{ij})^{-1}$ and
\begin{equation}
	\xi^\lambda_{ij}\left(f, \hat k\right)= \mathrm{e}^{\pi i f L_{ij} ( 1 - \hat{k}\cdot \hat l_{ij})}  \; \mathrm{sinc}\left(\pi f L_{ij} ( 1 + \hat{k}\cdot \hat l_{ij})\right) \; \frac{\hat l^a_{ij}\hat l^b_{ij}}{2}e^\lambda_{ab}(\hat k) \; ,
\end{equation}
where $\hat l_{ij} = (\vec x_j - \vec x_i)/|\vec x_j - \vec x_i|$ is a unit vector pointing from $i$ to $j$ and $e^\lambda_{ab}(\hat k)$ are the GW polarization tensors. Let us proceed by assuming that fluctuations in the arm lengths are negligible and the LISA configuration is perfectly equilateral at all times\footnote{In reality, several effects will contribute to breaking this perfectly symmetric configuration, leading to unequal and time-varying arm lengths. A more accurate description of the system should account for these modifications, which, e.g., will break the orthogonality of the usual AET Michelson TDI variables (see next section). For a discussion of some of these effects and their data analysis implications, see e.g. Ref.~\cite{Hartwig:2023pft}.}, so that $L_{ij} = L$ and $f_{ij} = f_*$. By substituting the statistical properties for a homogeneous, isotropic, and non-chiral SGWB GW signal
	\begin{equation} \label{eq:h-statistics}
		\langle \tilde h_\lambda(f,\hat k) \, \tilde h_{\lambda^\prime}^*(f', \hat k')\rangle = \delta(f - f')\delta(\hat k -\hat k')\delta_{ \lambda \lambda^{\prime} }\frac{P_{h}^{\lambda \lambda^{\prime}}(f)}{16\pi} \; \qquad \qquad  \langle \tilde h_\lambda(f,\hat k) \, \tilde h_{\lambda^\prime}(f',\hat k')\rangle = 0 \; ,
	\end{equation}
where $P_{h}^{\lambda \lambda^{\prime}}(f)$ denotes the one-sided GW PSD, and comparing with Eq.~\eqref{eq:data_correlation}, we express the signal PSD as
\begin{equation}
		S^{\mathrm{GW}}_{ij,mn}(f) \equiv \sum_\lambda \mathcal{R}^\lambda_{ij,mn} \, P_h^{\lambda \lambda}(f)  \equiv \left( \frac{f}{f_*} \right)^2 \sum_\lambda P_h^{\lambda \lambda }(f)  \; \int \frac{\textrm{d} \Omega_{\hat{k}}}{4 \pi}  \; \textrm{e}^{-2\pi i f \hat{k}\cdot (\vec x_i - \vec x_m)}  \;  \xi^\lambda_{ij}(f, \hat k)  \, \xi^\lambda_{mn}(f, \hat k)^* \; .
		\label{eq:CSDsignal}
\end{equation}
Here, $\mathcal{R}^\lambda_{ij,mn} $ are the (polarization-dependent) single link response functions.

As far as noise is concerned, we assume two contributions dominate the noise budget: Test Mass (TM) noise, typically dominating at low-frequency, and Optical Metrology System (OMS) noise, typically dominating at high frequencies. These are the two preeminent secondary noises that remain unsuppressed at the end of the TDI procedure\footnote{A more realistic noise model would also have to account for subdominant contributions, e.g., the tilt-to-length noise~\cite{Hartig:2022nxt, Hartig:2022htm, Paczkowski:2022nrt, George:2022pky, Hartig:2023ofu, Armano:2023fmc}, due to angular jitter in the readout system).} (see next section). TM and OMS noise contribute to the single link measurement as
	\begin{equation}
		\eta_{ij}^\mathrm{N}(t) = n^\text{OMS}_{ij}(t) +  D_{ij} n_{ji}^\text{TM}(t) + n_{ij}^\text{TM}(t) \; , \label{eq:eta}
	\end{equation}
where $D_{ij} $ denotes the time-delay operator that, under the assumptions of the present work (static and equilateral LISA constellation), acts on any function of time $x(t)$ as $D_{ij}x(t) = x(t - L)$, which, in the frequency domain, reduces to a phase shift represented by a multiplicative $\exp\{-2 \pi i f L \}$ factor. To express the noise contribution to Eq.~\eqref{eq:data_correlation}, we should then proceed by computing the noise PSD $S^{\rm N} (f)$. For this purpose, we assume the individual noise terms to be uncorrelated and zero mean so that
\begin{equation}
	\langle \tilde n^\text{TM}_{ij}(f) \, \tilde n^\text{TM*}_{lm}(f') \rangle = \frac{\delta_{ij,lm}}{2} \, S^\text{TM}(f) \, \delta(f - f') \;, \qquad 
   \langle \tilde n^\text{OMS}_{ij}(f) \, \tilde n^\text{OMS*}_{lm}(f') \rangle = \frac{\delta_{ij,lm}}{2}  \, S^\text{OMS}(f) \,\delta(f - f') \; ,
 \label{eq:csd_noise}
\end{equation}
This in turn assumes that all the TM and OMS components are equal. Furthermore, following Ref.~\cite{LISA:2017pwj}, the PSDs are given by:
\begin{equation}
	\begin{aligned}
			\label{eq:TM_OMS_noise_def}
			S^\text{TM}(f) &= A^2 \times 10^{-30} \;  \times \left[1 + \left(\frac{ 0.4 \textrm{mHz}}{f}\right)^2 \right]  \left[1 + \left(\frac{f}{ 8 \textrm{mHz}} \right)^4 \right] \times \left(\frac{1}{2 \pi f c} \right)^2 \; \times ( \textrm{m}^2 / \textrm{s}^3 ) \;,  \\ 
			S^\text{OMS}(f) &= P^2 \times 10^{-24} \;  \times \left[1 + \left(\frac{2 \times 10^{-3} \textrm{Hz} }{f} \right)^4 \right]  \times \left(\frac{2 \pi f}{c} \right)^2 \; \times ( \textrm{m}^2 / \textrm{Hz} ) \;.
	\end{aligned}
\end{equation}
In these expressions, the amplitudes of the TM and OMS noise PSDs are controlled by the dimensionless $A$ and $P$ parameters, respectively. To reproduce the noise level specified in~\cite{LISA_sciRD}, we set the fiducial values for these parameters to be $A=3$, $P=15$.

We stress that this is an overly simplified model of noise expected in LISA. For example, note that LPF's in-flight noise differs in level and shape from predictions, mostly at frequencies below $10^{-3}$ Hz. Moreover, we have not included any non-stationarity, such as glitches or drifts in the instrument noise. 
These effects will almost certainly be present in realistic data and will need to be taken into account in the development of data analysis pipelines.  
For longer observation times, possible anisotropies in the SGWB would also need to be taken into account.

\subsection{Projection on the TDI variables and likelihood.} 

\noindent TDI~\cite{Armstrong_1999, Tinto:1999yr, Estabrook:2000ef, Tinto:2020fcc, Prince:2002hp, Shaddock:2003dj, Tinto:2003vj} is a data processing technique consisting of combining several interferometric measurements, typically performed at different times, that will be used in LISA to suppress the primary noise sources (mostly laser noise, a white noise contribution several orders of magnitude larger than the required noise level). While several TDI variables can achieve the target noise suppression~\cite{Shaddock:2003bc, Vallisneri:2005ji, Muratore:2020mdf, Muratore:2021uqj}, in this work, we focus on the most commonly used variables, the Michelson XYZ variables, and their orthogonal combinations, typically referred to as AET TDI variables. Moreover, several generations of TDI variables exist, which achieve noise cancellation in scenarios with increasing complication and realism. In the present work, we employ first-generation TDI variables, which can suppress primary noises for a constellation with unequal (but constant) arm lengths. This choice is sufficient, given that we restrict ourselves to the case of a maximally symmetric, i.e., equilateral and equal noise levels, configuration. In this framework, the X TDI variable is defined as:
\begin{equation}
	{\rm X}  = (1 - D_{13}D_{31})(\eta_{12} + D_{12} \eta_{21}) + (D_{12}D_{21} - 1)(\eta_{13} + D_{13} \eta_{31}) \; ,
\end{equation}
and the Y and Z variables correspond to cyclic permutations of the three satellites. Effectively, working in the XYZ TDI basis consists of considering three interferometers that share their arms, leading to correlated measurements. For this reason, it is customary to introduce the AET basis, which, under the assumptions made throughout this work, can be shown to be orthogonal. For explicit expressions of the noise PSDs in the AET basis see, e.g. Ref.~\cite{Flauger:2020qyi, Hartwig:2023pft}. 

We recall that for simplicity, in the analysis carried out in this work, we used the X TDI variable only, the generalization to all channels and to higher generation TDI variables is left for future work. In particular, exploiting the different response functions of these channels will enable some discrimination between signal and noise~\cite{Flauger:2020qyi}, although degeneracies remain when the instrument is modelled more realistically. 
Under the assumptions of this work, a generalization to three TDI channels is straightforward, however relaxing some assumptions on the symmetry of the configuration would require more elaborate treatment~\cite{Hartwig:2023pft}.

As discussed in Ref.~\cite{Flauger:2020qyi}, the data obtained following the procedure discussed in the main text presents mild non-Gaussianity. Thus, the (log-)likelihood should include some skewness corrections to model this component and avoid a systematic bias in the results. It is known~\cite{Bond:1998qg, Sievers:2002tq, WMAP:2003pyh, Hamimeche:2008ai} that an appropriate (log-)likelihood has form:
	\begin{equation}
		\ln \mathcal{L} (\vec{\theta} | D_k) = \frac{1}{3} \ln \mathcal{L}_{\rm G} (\vec{\theta} | D_k) +  \frac{2}{3} \ln \mathcal{L}_{\rm LN} (\vec{\theta} | D_k) \; ,
	\end{equation}
where $\vec{\theta} = \{\vec{\theta}_s, \vec{\theta}_n\} $ are the parameters (with $\vec{\theta}_s$, $\vec{\theta}_n$ being the signal and noise parameters, respectively), and $\ln \mathcal{L}_{\rm G}$, $\ln \mathcal{L}_{\rm LN}$ are a Gaussian and log-normal likelihood: 
\begin{equation}
		\ln \mathcal{L}_{\rm G} (\vec{\theta} | D_k)  = -\frac{N_d}{2} \sum_{k} w_k \left[ 1 - D_k / \mathcal{D}_{k} (\vec{\theta}) \right]^2  \; , \qquad 
  \ln \mathcal{L}_{\rm LN} (\vec{\theta} | D_k )  = -\frac{N_d}{2} \sum_{k}  w_k \ln^2 \left[ \mathcal{D}_{k}  (\vec{\theta}) / D_k   \right] \; .
\end{equation}
Here, $\mathcal{D}_{k}(\boldsymbol{\theta})$ denotes the theoretical model for the data $ \mathcal{D}_{k} (\boldsymbol{\theta}) = \Omega_{  {\rm GW} } (f_k, \boldsymbol{\theta}) + \Omega_{ {\rm n} } (f_k, \boldsymbol{\theta})$ , with $\Omega_{ {\rm GW} } (f_k, \boldsymbol{\theta})$ and $\Omega_{{\rm n}} (f_k, \boldsymbol{\theta})$, being the signal and noise model, respectively.

\section{Technical details, coverage tests and performance}\label{app:tmnre}

\noindent In this section, we report some technical details concerning the implementation of our technique, discuss its computational performance, and present coverage test results for the case study \textbf{C4}, which includes the additional transient sources. Starting with technical details regarding the implementation of \texttt{saqqara}, several numerical settings should be chosen for the general structure of the algorithm, as well as the network architecture. Each of the options is explained in detail within the configuration files in the \texttt{saqqara} repository, however, here we detail the choices made to produce the results in this work. Firstly, in each round of inference, we use 500,000 simulations. For training the network, we train for a maximum of $50$ epochs, and set the patience before early stopping is triggered (due to a non-decreasing validation loss) to $7$ epochs. Note that after training, we then reset the network to the state with the lowest validation loss. We take training and validation batch sizes of $512$, splitting the simulation dataset in the ratio $0.9:0.1$, and start with an initial learning rate of $2 \times 10^{-5}$. In terms of inference, for truncating the 1-dimensional priors, we take $\alpha = 10^{-5}$ (in the sense of Refs.~\cite{Miller:2021hys,Bhardwaj:2023xph}). For the sampler, we follow the notation in Ref.~\cite{AnauMontel:2023stj} and take $\epsilon = 10^{-3}$, $\log \mathcal{L}_\mathrm{max} = 500$, $n_\mathrm{batch} = 10$, $n_\mathrm{samples/slice} = 20$, and $n_\mathrm{steps} = 4$. As mentioned, these are detailed both in the \texttt{saqqara} repository, as well as in Ref.~\cite{AnauMontel:2023stj}. Finally, for the MCMC runs, we take $n_\mathrm{burn} = 500$, $n_\mathrm{steps} = 1000$, $r_\mathrm{conv} = 10^{-3}$, and a maximum of $50$ iterations.

\begin{figure*}[t]
    \centering
    \includegraphics[width=\linewidth]{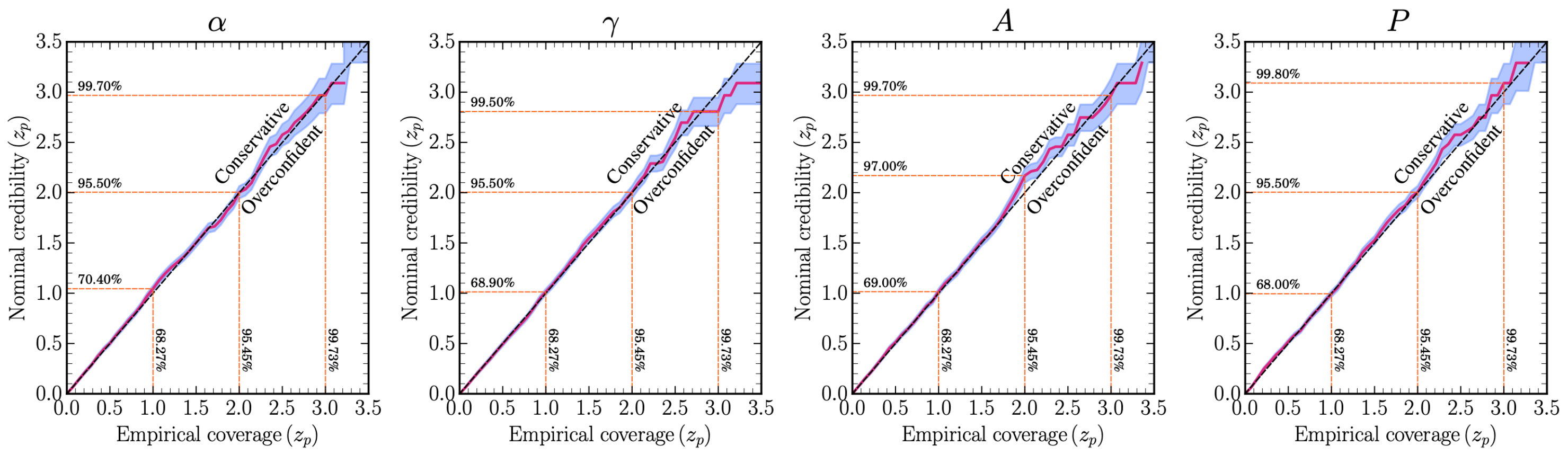}
    \caption{Coverage test results for the analysis of the power law template in the presence of additional transient signals (case study \textbf{C4}). From \emph{left} to \emph{right}, the panels show the coverage statistics for each of the four parameters in the model ($\alpha$, $\gamma$, $A$, $P$). For all panels, the empirical coverage is shown on the horizontal axis, and the nominal coverage is shown on the vertical axis. The 68\% confidence interval on the coverage is shown by the blue shaded region, and the central value is shown by the pink line.}\vspace*{-15pt}
    \label{fig:coverage-supp}
\end{figure*}

As mentioned in the main text, coverage tests are an important consistency check of our SBI pipeline, especially in scenarios where a comparison to a traditional method like MCMC is not possible (either computationally, or statistically). As a brief review, coverage tests (see e.g. Ref.~\cite{Hermans:2021aaa} for a more complete discussion in the context of SBI) implement the idea that in Bayesian parameter estimation, repeated inference over different noise/statistical realisations of the same signal should result in posteriors that shift relative to the true value. The rationale behind this sort of expected coverage test is that -- simply as a result of statistical fluctuations -- the $x$\% credible interval for the posterior should contain the simulation-truth value $x$\% of the time. To carry out this test for the case study in this work, we generated 1000 additional test simulations generated from the truncated prior. We then perform inference on each observation. In each case we can note how often the injected value was contained inside the $x$\% confidence interval and construct a cumulative distribution. A well calibrated posterior distribution will be a totally diagonal line when the expected coverage is plotted against the empirical findings. The results for case study \textbf{C4} are shown in Fig.~\ref{fig:coverage-supp} for each model parameter. We see that in every case, we obtain extremely well calibrated coverage for our posterior estimates. This strongly supports our claim in the main text of recovering unbiased posteriors for noise and SGWB parameters even in the presence of transient signals.

The final relevant discussion point concerns the computational performance of our SBI algorithm. In terms of computational complexity, there are a number of steps to generating posteriors; simulation/data generation (which is fully parallelised within \texttt{saqqara}); network training/likelihood-to-evidence estimation; and inference. With the setup considered here, we perform the inference in two steps. The first of these learns the individual marginal posteriors for the SGWB and noise parameters. This step is fully amortised (in the sense that once the training is complete, the inference is almost immediate on any signal) and allows us to efficiently ``zoom in" (or truncate) to the prior region most relevant to the given observation, see Refs.~\cite{Miller:2021hys,Bhardwaj:2023xph} for more details on this process. Then, in the second step, we use the techniques developed in Ref.~\cite{AnauMontel:2023stj} to estimate the full joint posterior, which we explore with a \texttt{pytorch}-based sampling technique.\footnote{We could also have estimated, e.g., 1- or 2-dimensional marginals for any/all parameters, depending on the specific inference needs.} In terms of timing (on a 20 CPU-core resource with a single \emph{NVIDIA GeForce GT 73} graphics card), the 500,000 simulations we use in each step took around 15 minutes to generate, and the subsequent network training took an additional 25 minutes (which does not need to be repeated). For the 1-dimensional ratio estimators, the inference is then essentially instantaneous, only requiring a single network evaluation. For the higher-dimensional marginals, the sampling adds a slight overhead (around 7 minutes in e.g. case study \textbf{C1}).

\section{Full results for agnostic template fits}\label{app:agnostic}

\begin{figure*}[htb]
    \centering
    \includegraphics[width=\linewidth]{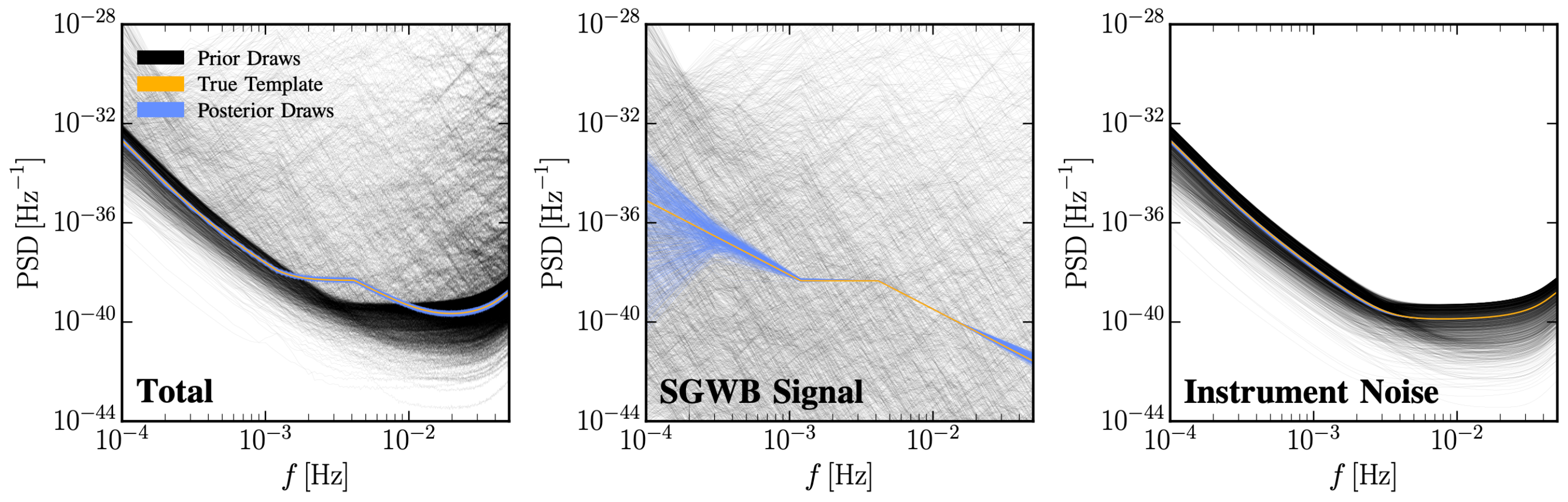}
    \caption{Reconstruction of SGWB and instrumental noise components for an agnostic template with five signal bins. In all panels, draws from the initial Bayesian prior are shown in black, whereas the resulting posterior draws are shown in blue. The injected signal is shown in yellow. \textit{Left panel.} Reconstruction of the total power-spectral density. \textit{Middle panel.} Reconstruction of the individual SGWB component. \textit{Right panel.} Reconstruction of the instrumental noise contribution.}
    \label{fig:agnostic5b-supp}
\end{figure*}

\begin{figure*}[htb]
    \centering
    \includegraphics[width=\linewidth]{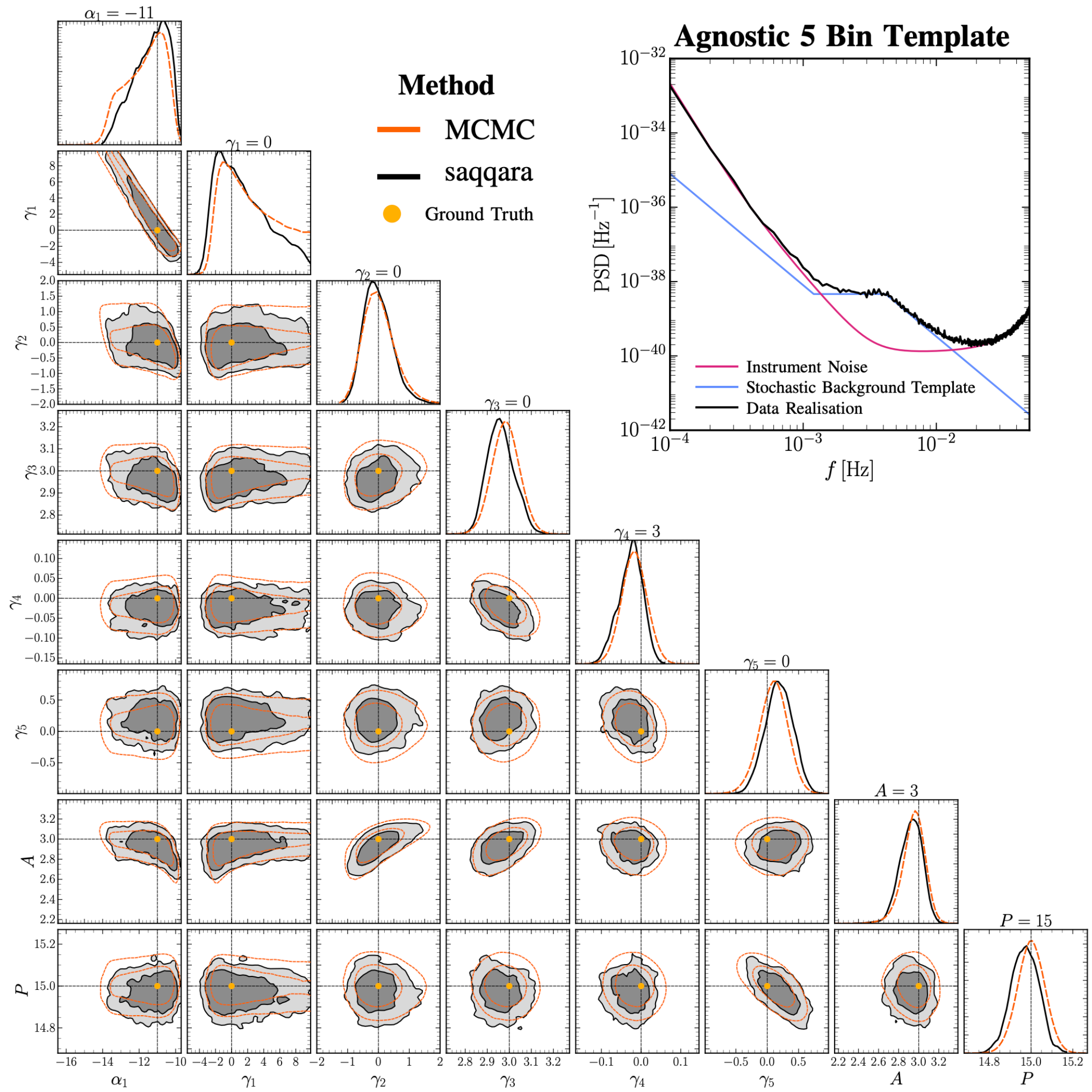}
    \caption{Parameter constraints for agnostic template fit with five signal bins. \textit{Main plot.} Corner plot highlighting the relative agreement between the MCMC (shown by orange dashed lines) and SBI (shown in solid black) approaches for all parameters in the agnostic template fit with five signal bins. The true injected values are highlighted by the dashed black horizontal and vertical lines, and by the yellow markers. \textit{Upper inset.} Illustration of the explicit data realisation (black line) for the case study \textbf{C2} along with the injected instrumental noise (pink line) and stochastic background signal (blue line).}
    \label{fig:agnostic5bR-supp}
\end{figure*}

\begin{figure*}[htb]
    \centering
    \includegraphics[width=\linewidth]{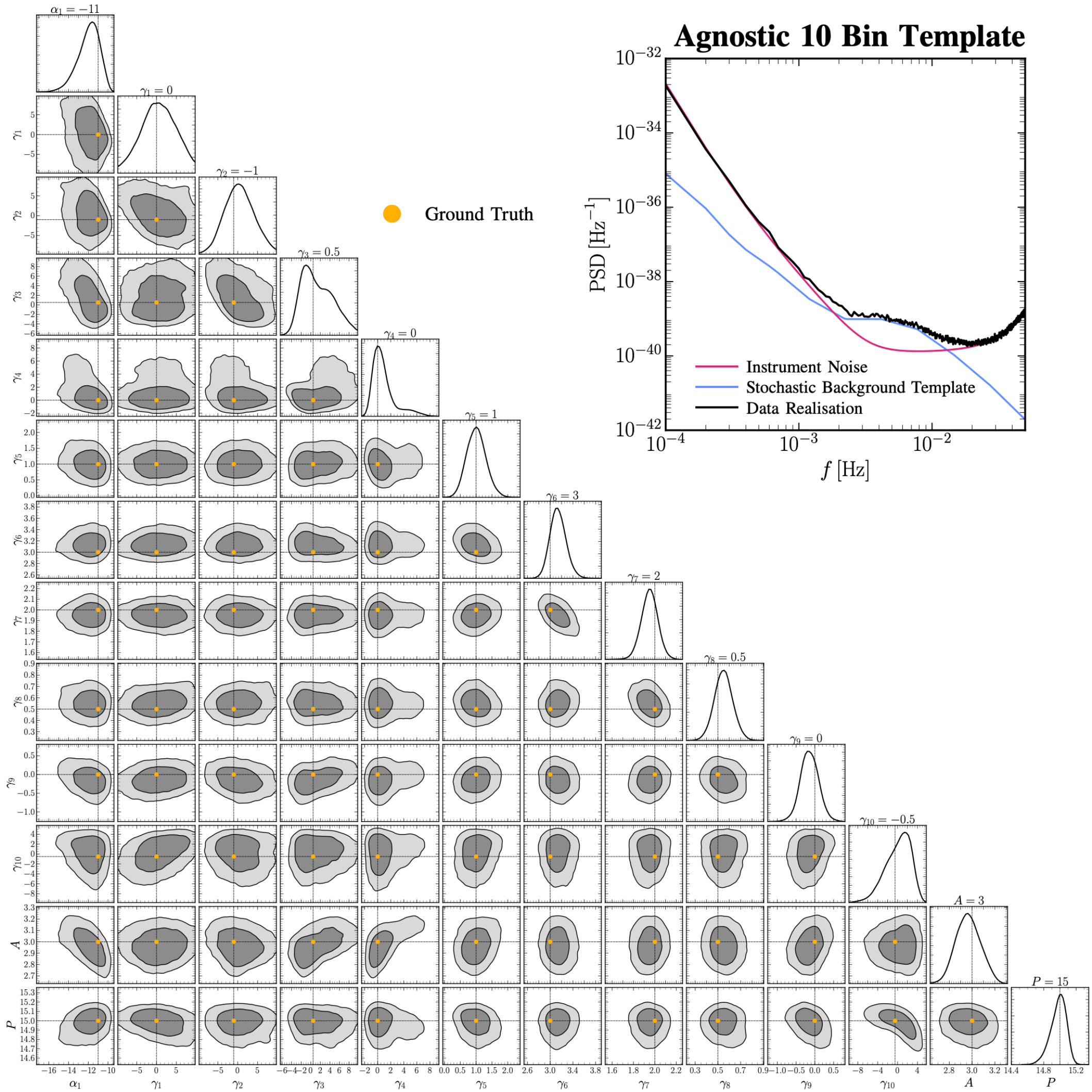}
    \caption{Parameter constraints for agnostic template fit with five signal bins. \textit{Main plot.} Corner plot highlighting the parameter constraints using the SBI approaches for all parameters in the agnostic template fit with ten signal bins. The true injected values are highlighted by the dashed black horizontal and vertical lines, and by the yellow markers. \textit{Upper inset.} Illustration of the explicit data realisation (black line) for the case study \textbf{C3} along with the injected instrumental noise (pink line) and stochastic background signal (blue line).}
    \label{fig:agnostic10bR-supp}
\end{figure*}

\noindent In addition to the results presented in the main text, here we show additional results for the case studies \textbf{C2} and \textbf{C3}. In particular, in Fig.~\ref{fig:agnostic5b-supp}, we illustrate the reconstruction of the stochastic background and instrumental contributions in the context of the agnostic template with ten signal bins. Furthermore, in Fig.~\ref{fig:agnostic5bR-supp}, we present the analysis at the level of parameter constraints for the agnostic template fit with five signal bins. We also show the comparison with the MCMC approach, which we see agrees extremely well with our SBI approach. Finally, in Fig.~\ref{fig:agnostic10bR-supp}, we show the corresponding analysis for the ten bin agnostic fit.

\section{Review of Simulation-based Inference and TMNRE}

\noindent In this section we provide a brief review of simulation-based inference. Specifically, we cover the various classes of SBI methods before focussing on the technicalities of TMNRE.

\subsection{Classes of SBI Algorithm}

\noindent All SBI algorithms are designed to answer the same question: \emph{how can we do robust Bayesian inference given an implicit representation of the likelihood through a generative model?} In other words, suppose we are only given a ``simulator" that takes model parameters $\theta = (\theta_1, \theta_2, \ldots)$ and stochastically generates data $x$, then the key idea is that running this simulator is equivalent to sampling from the likelihood $x \sim p(x | \theta)$. This is the origin of the term ``likelihood-free inference", although this has now been replaced by the more appropriate ``simulation-based" or ``implicit likelihood" description~\cite{Cranmer:2019eaq}. It is worth noting that many of the recent advances in SBI have been facilitated by corresponding developments in machine learning. This has opened up the opportunity to use SBI methods to analyse high-dimensional and complex data structures such as images, time series, point clouds etc. On a more historical note, all of these algorithms move beyond the paradigm of Approximate Bayesian Computation (ABC)~\cite{Sunnaker:2013aaa}, which requires the choice of a hand-crafted summary statistic and distance measure to quantify the similarity between two sets of data.

There are a number of classes of SBI algorithms that vary in terms of how they estimate the relevant quantities in Bayes' theorem. In each case, the general goal is to obtain the posterior $p(\theta | x) = p(x | \theta) p(\theta) / p(x)$, where $p(\theta)$ is the (Bayesian) prior and $p(x)$ is the (Bayesian) evidence. In particular, they can be broadly classified as follows:

\begin{itemize}
    \item \emph{Neural Posterior Estimation (NPE).} These methods aim to directly estimate the posterior $p(\theta | x)$, typically utilising neural network structures such as normalising flows, which are manifestly normalised probability densities and easy to sample from~\cite{Papamakarios:2016aaa}. This has been used successfully in a number of contexts including compact binary GW data analysis~\cite{Dax:2021tsq,Wildberger:2022agw,Dax:2022pxd}.
    \item \emph{Neural Likelihood Estimation (NLE).} In contrast to NPE, likelihood-estimation techniques construct an approximation to the (simulated) data likelihood $p(x | \theta)$~\cite{Papamakarios:2016aaa,Alsing:2019xrx,Lin:2022ayr}. This can then be sampled using traditional stochastic sampling techniques such as MCMC or nested sampling.
    \item \emph{Neural Ratio Estimation (NRE).} The third class of methods is neural ratio estimation~\cite{Miller:2021hys,Rozet:2022aaa,Delaunoy:2022aaa,Miller:2022haf,Hermans:2019aaa,Durkan:2020aaa}, which, contrary to the two mentioned above, approaches the Bayesian inference problem by constructing an estimate of the likelihood-to-evidence ratio $r(x, \vartheta) = p(x | \vartheta) / p(x) = p(\vartheta | x)/p(\vartheta)$, where $\vartheta$ is some collection of parameters in $\theta$, or derived parameters. The equality here follows simply as a result of probability rules and therefore if one can construct $r(x, \vartheta)$, then it is possible to directly access the posterior by re-weighting prior samples. This work is based on a specific implementation of NRE, \textbf{truncated marginal neural ratio estimation} (TMNRE)~\cite{Miller:2022shs}, which we discuss in detail below.
\end{itemize}

\subsection{Truncated Marginal Neural Ratio Estimation (TMNRE)}

\noindent In the main text, we discussed the reasons why we believe that, from the big-picture perspective, TMNRE is a suitable algorithm for the application of SBI to SGWB analysis. Broadly, these were focussed on two key properties of the algorithm: the truncation/sequential aspect (the `T'), and the ability to directly estimate marginal posteriors (the `M'). We argued that we expected these to lead to significant simulation efficiency and statistical flexibility when applied to the LISA data analysis challenge. In this section, we briefly review the technical aspects of our TMNRE algorithm. It is also worth noting that TMNRE has been successfully applied in a number of scenarios beyond SGWB and LISA data analysis, including CMB and 21cm cosmology, stellar streams, strong lensing image analysis, point sources, and GWs from compact binaries~\cite{Alvey:2023naa,Bhardwaj:2023xph,Alvey:2023pkx,Saxena:2023tue,Gagnon-Hartman:2023soa,Coogan:2022cky,Montel:2022fhv,Cole:2021gwr,AnauMontel:2022ppb}.

\vspace{10pt}
\noindent \emph{Estimating the ratio $r(x, \vartheta)$.} The first point to clarify about the TMNRE algorithm is how it estimates $r(x, \vartheta)$. In particular, we note here that  that $\vartheta$ need not be the full set of model parameters $\theta$. In what follows, it could be a single parameter $\vartheta = \theta_i$, a set of parameters $\vartheta = (\theta_i, \theta_j, \ldots)$, or some derived model parameter. This directly encodes the marginal aspect of the algorithm and defines what we mean in the main text with ``implicitly marginalising" over, \emph{e.g.}, the parameters of transient signals. With this clarified, the estimation of the ratio proceeds as follows: firstly, we note that we can re-express $r(x, \vartheta) = p(x | \vartheta)/p(x) = p(x, \vartheta) / p(x)p(\vartheta)$. In other words, $r(x, \vartheta)$ is the ratio between a \emph{joint} sample $x, \vartheta \sim p(x, \vartheta) = p(x | \vartheta) p(\vartheta)$ and a \emph{marginal} sample $x, \vartheta \sim p(x) p(\vartheta)$. It is worth noting that for any choice of $\vartheta$, it is trivial to get either set of samples. Joint samples $x, \vartheta \sim p(x | \vartheta) p(\vartheta)$, properly marginalised over the variation in the other parameters, are obtained by running the simulator on prior samples from the full model $\theta \sim p(\theta)$. Parts of the parameter space that are not of interest can be simply discarded or masked. Similarly, to generate marginal samples, one can take a pair $(x, \theta)$ from a simulation run, and re-sample $\theta \sim p(\theta)$. Again irrelevant parameters can be discarded. 

The second step for ratio estimation is to construct a binary classification task between joint and marginal samples. Specifically, the goal is to find a classifier $d_\phi(x, \vartheta)$ with some trainable parameters $\phi$ (almost always in the form of a neural network and its weights) that optimally outputs, \emph{e.g.}, $d_\phi(x, \vartheta) = 0$ if $x, \vartheta$ is a joint sample and $d_\phi(x, \vartheta) = 1$ if it is drawn marginally. Said differently, we can rephrase the task of performing parameter inference as a classification (marginal vs. joint samples) problem, which is extremely well suited to modern supervised machine learning techniques. In practice, the mapping of ratio estimation onto the binary classification task defined above is realized using the standard TMNRE (binary cross-entropy) loss function~\cite{Miller:2022shs},
\begin{equation}
\mathcal{L}[f_\phi] = - \int{\mathrm{d}x\,\mathrm{d}\vartheta \, \left[p(x, \vartheta) \, \ln (\sigma (f_\phi(x, \vartheta))) + p(x)p(\vartheta) \, \ln (1 - \sigma (f_\phi(x, \vartheta))) \right]},
\end{equation}
where $d_\phi(x, \vartheta) = \sigma(f_\phi(x, \vartheta))$ and $\sigma(x) = (1 + e^{-x})^{-1}$ is the sigmoid function. The main motivation to justify such choice for the loss function is that it can be shown analytically (\emph{i.e.}, taking a functional derivative with respect to $f_\phi$) that the \emph{optimal} classifier $f_\phi^\star(x, \vartheta)$ is precisely $f_\phi^\star(x, \vartheta) = \ln r(x, \vartheta)$. 

The above review explains the general reason that TMNRE works as a parameter inference algorithm, and also highlights how it can be used to implicitly marginalise over any part of the model. To further emphasise this point, TMNRE does not require any hand-crafted or analytic method for marginalisation, just the ability to sample from a simulator, and mask/ignore the parameters that you wish to marginalise over.

\vspace{10pt}
\noindent \emph{Truncation.} Before we end this section by discussing the relevant algorithm design choices in the context of SGWB analysis, we will briefly explain how truncation, or ``zooming-in" is achieved. To illustrate this process, imagine that $\vartheta$ is just a single parameter $\vartheta = \theta_i$ and that we have a learned estimator $\hat{r}(x, \theta_i)$ of the ratio $r(x, \theta_i)$. Then, to obtain posterior estimates for some observation $x_0$, we can sample from the prior $\theta_i \sim p(\theta_i)$ and re-weight by the ratio $r(x_0, \theta_i) = p(\theta_i | x_0)/p(\theta_i)$. In this process, if $\theta_i$ is a well-measured parameter, then there will be regions where $r(x_0, \theta_i)$ is very small. The general intuition behind the truncation part of TMNRE is to sequentially exclude these regions from the initial prior range (without changing the shape of the prior), and re-simulate new data targeting the region of interest. This approach has been shown to be extremely simulation efficient compared to either traditional joint inference or amortised techniques~\cite{Cole:2021gwr,Bhardwaj:2023xph,Alvey:2023naa} to analyse data for individual GW observations. See Refs.~\cite{Miller:2022shs,AnauMontel:2023stj} for a technical description of how this is achieved in single- and multi-parameter setups, as well as the various precision settings that are required to achieve efficient but conservative truncation.

\subsection{Design Choice for SGWB Analysis}

\noindent Within this general framework, there are a number of design choices that are relevant to the application of TMNRE to SGWB analysis. The most obvious of these is the forward simulation model, which we discuss in the main text. The second set of choices are the TMNRE settings which are also discussed above. The final concrete design specifications that are required concern the design of the network architecture for the classifier $f_\phi(x, \vartheta)$. As we mention in the main text, for the application of TMNRE, this network typically splits into two components: a compression network $\tilde{s}(x)$ and a ratio estimator $\tilde{r}(s, \vartheta)$, which are combined to get $f_\phi(x, \vartheta) = \tilde{r}(\tilde{s}(x), \vartheta)$. Importantly, both the compression network $\tilde{s}$ and $\tilde{r}$ are trained simultaneously. This ensures that no hand-crafted compression statistics are required, but rather, they are learnt directly. In our implementation, we use a relatively standard form for the ratio estimator $\tilde{r}(s, \vartheta)$, see e.g. Refs.~\cite{Miller:2022shs,AnauMontel:2023stj}, and therefore we spend the rest of the section explaining the choices for the compression network $\tilde{s}(x)$.

Ultimately the compression network architecture that we chose for $\tilde{s}(x)$ is motivated by the data that we are trying to analyse for extracting the SGWB. In particular, in terms of structure, the data $x$ consists of noise variances $\langle \bar{d}^\star(f_i) \bar{d}(f_i) \rangle$ across a sequence of frequency bins $f_i$, in one or more channels. We know that broadly, the signal we are looking for in this work is characterised by a set of parameters $\theta_\mathrm{SGWB}$ that define a template that spans the various bins. Beyond this, we know that there are/could be additional contributions from instrumental noise or transient sources that may induce excesses to the signal, or correlated, non-Gaussian statistics across frequency bins. This physical intuition motivated the main component of the compression network in \texttt{saqqara}, which is essentially a 1-dimensional version of a \texttt{unet} architecture, similar to that described in Ref.~\cite{Bhardwaj:2023xph}. In simple terms, the \texttt{unet} consists of two parts: a part that shrinks the initial data down, and a second part that rescales it back up. In the downscaling part, the goal is to gradually downsample (using a sequence of convolutional and max-pooling layers) the data and extract sequentially more fine-grained features. This creates an information bottleneck at the bottom of the `U' structure which encourages the network to extract the most important features from the data. In the decompression step, the \texttt{unet} attempts to rebuild the data, with various segments identified and classified. The final technical addition to the network architecture is the existence of `skip` connections, which allow for the higher-level features learned in the downsampling steps to be used in the corresponding reconstruction. In the context of LISA data, we can imagine this architecture first extracting the relevant frequency bins for a given SGWB signal, before focussing on the fine-grained details that are controlled by e.g. the template, and then reconstructing the signal. These are very general statements of course, but one of our aims with the \texttt{saqqara} (see \href{https://github.com/PEREGRINE-GW/saqqara}{github} repository) pipeline is to be agnostic to the classes of excess on top of an SGWB signal.

\vfill
\vfill

\end{document}